\documentclass[
aps,prl,amsmath,amssymb,amsfonts,showpacs,noshowkeys,
groupedaddress,a4paper,reprint]{revtex4-1}
 \usepackage{graphicx}
 \usepackage{bm}
 \usepackage[version=3]{mhchem}
 \usepackage[english]{babel}
 \usepackage[T1]{fontenc}
 \usepackage{textcomp}
 \usepackage{xcolor}
 \usepackage{hyperref}

\begin{document}
%
%
\title{Correlated Particle Motion and THz Spectral Response %
of Supercritical Water}
\author{Maciej \'Smiechowski}
\email[Corresponding author. Electronic mail:\\]{Maciej.Smiechowski@pg.gda.pl}
\affiliation{
Department of Physical Chemistry, Chemical Faculty,
Gda\'nsk University of Technology,
Narutowicza 11/12, 80-233 Gda\'nsk, Poland}
\affiliation{
Lehrstuhl f\"ur Theoretische Chemie,
Ruhr-Universit\"at Bochum,
44780 Bochum, Germany}
\author{Christoph Schran}
\author{Harald Forbert}
\author{Dominik Marx}
\affiliation{
Lehrstuhl f\"ur Theoretische Chemie,
Ruhr-Universit\"at Bochum,
44780 Bochum, Germany}
\date{\today}
\frenchspacing

\begin{abstract}
Molecular dynamics simulations of supercritical water
reveal distinctly different distance-dependent modulations of
dipolar response
and correlations in particle motion compared to ambient conditions.
The strongly perturbed
H-bond network of water at supercritical conditions
allows for considerable translational and rotational freedom of individual
molecules.
These changes give rise to substantially different
infrared spectra and vibrational density of states
at THz frequencies for densities above and below the Widom line
that separates percolating liquid-like and clustered gas-like
supercritical water.
\end{abstract}


\maketitle

Supercritical matter is an ideal testing ground to systematically
study fundamental aspects of fluids as a function of density
\cite{Cunsolo1998,Gorelli2006,Simeoni2010,McMillan2010,Bolmatov2013,%
Gallo2014,Sokhan2015,Fomin2015}.
Although formally being single phase systems,
supercritical fluids nevertheless show two distinct
regions that are separated by the so-called Widom line,
distinguishing gas-like and liquid-like dynamical regimes that are
reminiscent of the respective subcritical domains
\cite{Simeoni2010,McMillan2010,Gallo2014,Sokhan2015,Fomin2015}.
Hence,
supercritical fluids are not only far more complex in nature than
believed for a long time, but the distinct regimes also encompass
vastly different properties.
Transcending fundamentals, the study of liquids above their
critical point (CP) is also motivated by their increasing importance for
large-scale industrial processes, serving e.g. as environmentally
friendly ``green solvents'' \cite{Li2008,DeSimone2002}.

Most notably supercritical water (SCW; CP:
$T_{\rm c}=647$~K, $p_{\rm c}=22.1$~MPa, $\rho_{\rm c}=0.32$~g/cm$^{3}$)
becomes an increasingly important processing medium for key transformations
in benign aqueous environments \cite{Jessop1995}.
The latter owes much to the dramatically reduced dielectric constant of SCW
(from $\sim 80$ at ambient conditions to $\sim 6$ just above the CP)
\cite{Pan2013}.
On the other hand, SCW becomes fairly conducting
up to the point of metalizing above 7000~K in the GPa~pressure
range \cite{Cavazzoni1999}.
All this makes SCW a ``tunable fluid environment''
with amazing properties \cite{Weingartner2005}.
The macroscopic phenomena that underlie these peculiarities at the microscopic
level seem to be driven chiefly by significant changes in the
hydrogen bond (H-bond) network of water at extreme conditions.
Such structure of SCW is well-studied using neutron/x-ray diffraction (ND/XRD)
\cite{Postorino1993,Ohtaki1997,Bernabei2008b},
quasi-elastic neutron scattering (QENS)~\cite{Tassaing2000}
with energy transfers up to 100~meV ($\approx 800$~cm$^{-1}$)
and
deep inelastic neutron (Compton)/x-ray (Raman)
scattering (DINS/IXS) \cite{Sit2007,Pantalei2008,Sahle2013} in the eV~range,
thus spanning many orders of magnitude in time but also
in length scales depending on momentum transfer.
For instance, the dynamic structure factor obtained from incoherent QENS
of SCW~\cite{Tassaing2000} yields, in the limit $Q<1$~{\AA}$^{-1}$,
the density of states of H~atoms up to ca.~800~cm$^{-1}$
($\approx 24$~THz).
Complementing structure, dynamic relaxation~\cite{Matubayasi1997a}
of SCW has been probed as well by NMR.

Experiments on SCW can be interpreted
at the molecular level by analyzing molecular dynamics (MD) simulations
that can be directly compared with experimental data \cite{Kalinichev2001}.
The electronic structure of SCW is accessible via ab initio MD (AIMD)
\cite{Marx2009} that allow one to compute dipole moments
and vibrational spectra \cite{Boero2000} as well as
IXS~spectra \cite{Sahle2013}.
Moreover, AIMD simulations also provide computational access to
reactions in SCW \cite{Boero2003}.

When it comes to elucidating H-bond dynamics,
infrared (IR)~spectroscopy remains
the prime technique to study water \cite{Yagasaki2013}
including SCW \cite{Tassaing2002,Kandratsenka2008}.
IR~spectroscopy is sensitive primarily to atomic and/or molecular
motion that gives rise to dipole moment changes
along the associated oscillatory displacements~\cite{Handbook2006}.
Upon lowering the frequency, i.e. the excitation energy,
one can successively probe fast intramolecular vibrations
in the 4000--500~cm$^{-1}$ (120--15~THz) mid-IR
range, intermolecular dynamics such as H-bond motion or
hindered molecular rotations and translations
in the 500--50~cm$^{-1}$ (15--1.5~THz)
far-IR regime \cite{Handbook2006}, and
even slower phenomena such as dielectric relaxation
at GHz~frequencies~\cite{Ronne1999},
thus allowing one to probe specific processes on their intrinsic timescales.
Vibrational spectroscopy of water in the {\em mid-IR regime}
is particularly sensitive to cluster size, H-bond abundance and
H-bond strength as the {\em intramolecular} stretching
band of \ce{H2O} is the traditional probe of these
\cite{Buck2000,Tassaing2002,Kandratsenka2008}.
In recent years, however, the {\em far-IR region} of the vibrational spectrum,
now commonly referred to as the ``terahertz'' (THz) regime
(note that 100~cm${}^{-1} \approx 3$~THz $\approx$~12~meV),
has been appreciated to offer direct insights into the picosecond
dynamics~--- and thus into {\em intermolecular} motion
such as H-bond network dynamics being characteristic to liquid water.
Advances in THz~laser spectroscopy closed the ``THz gap''
in IR~spectroscopy
and now permit unprecedented insights for instance into the
modified H-bond dynamics in the vicinity of
molecules \cite{Heugen06,Ebbinghaus07,Tielrooij2009}.
For ambient water, it has been shown theoretically
that THz~spectroscopy indeed probes specific intermolecular modes
of the H-bond network \cite{Heyden2010}.
For aqueous solutions of simple ions,
the THz~response due to these perturbations has been computed
\cite{Smiechowski2013,Smiechowski2015}
and THz~peaks could even be assigned to distinct intermolecular modes
in the case of molecules \cite{Sun2014}.

In this Letter we investigate~--- with a clear focus on changes of the
THz~spectral response~--- the properties of water at two supercritical state
points that represent gas-like and liquid-like H-bond dynamics
in the spirit of Widom crossover.
We uncover to what extent the
dramatic changes in the dynamical H-bond network fluctuations
associated with crossing the percolation line
affect cross-correlations in particle motion in a spatially-resolved
frequency-domain picture, and how the intermolecular dipolar correlations
of supercritical fluids qualitatively change their
THz~responses compared to ambient water.

\begin{figure}
\includegraphics{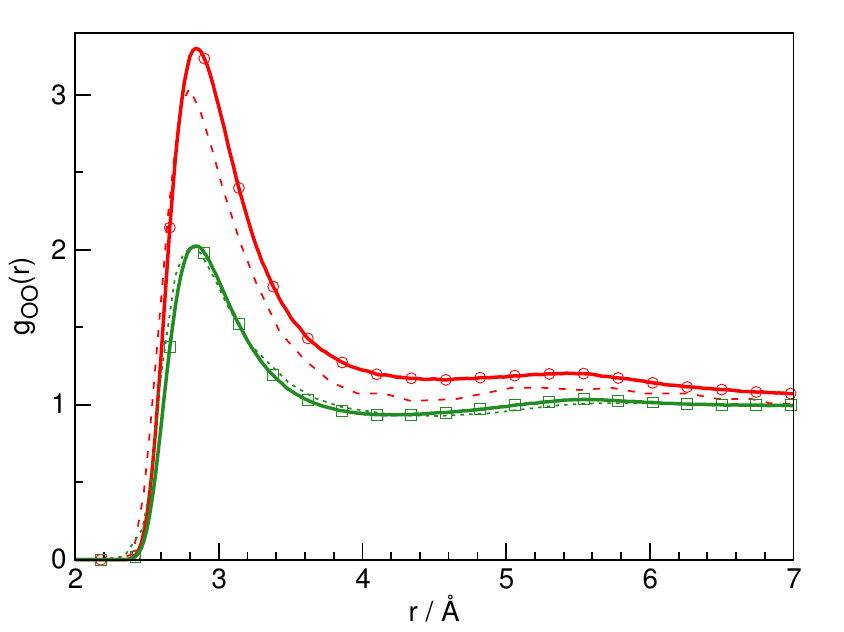}
\caption{\label{fig:rdfs}
Radial distribution functions $g_\mathrm{OO}(r)$ for LD--SCW (red line with
circles) and HD--SCW (green line with squares) compared to experimental data
at $\rho=0.23$~g/cm$^{3}$ (Ref.~\onlinecite{Funel1997}, dashed line)
and $0.58$~g/cm$^{3}$ (Ref.~\onlinecite{Bernabei2008b}, dotted line).}
\end{figure}

For this study we selected two supercritical state points (at $T=660$~K)
located deep in the two regimes separated by the Widom line \cite{Gallo2014}
(and by the percolation line of SCW \cite{Bernabei2008b}),
namely at $\rho=0.2$~g/cm$^{3}$
(low-density, LD--SCW) and at $0.6$~g/cm$^{3}$ (high-density, HD--SCW).
For reference, we also performed ambient condition
simulations at $300$~K and $0.997$~g/cm$^{3}$
(room-temperature water, RTW).
It is noted in passing that we must resort to force field MD
(see Supplemental Material, SM
\footnote{See Supplemental Material,
which includes Refs.~\onlinecite{Toukan1985,Lyubartsev2000,Raabe2007,Luzar1996,%
Kumar2007,Ramirez2004,Ivanov2013,Heyden2012,Jonchiere2011,Skaf2000,%
Bursulaya1999b}, for additional data, analyses, and computational details})
\nocite{Toukan1985,Lyubartsev2000,Raabe2007,Luzar1996,Kumar2007,Ramirez2004,%
Ivanov2013,Heyden2012,Jonchiere2011,Skaf2000,Bursulaya1999b}
in order to provide the statistical data base that allows
us to spatially decompose the long-ranged
intermolecular dependencies at THz~frequencies.
As a first test of the adequacy of these simulations, we investigate
the global structure of SCW as revealed by $g_\mathrm{OO}(r)$ in
Fig.~\ref{fig:rdfs}.
These radial distribution functions (RDFs) compare most favorably with the
available experimental data at matching densities (0.23~g/cm$^{3}$
\cite{Funel1997} and 0.58~g/cm$^{3}$ \cite{Bernabei2008b})
as do the corresponding O--H and H--H correlations~\cite{Note1}.

The corresponding microscopic structure
is reflected in the H-bonding properties of the fluids.
In contrast to RTW, where essentially all molecules form an inter-connected
three-dimensional H-bond network, SCW is characterized by the appearance of
isolated patches (``clusters'') that fluctuate rapidly on a
\mbox{(sub-)}picosecond time scale \cite{Churakov1999}.
The lifetime of these clusters is of course somewhat dependent on the
applied H-bond definition \cite{Kalinichev1994}
and we follow here a variant of the standard geometric criterion \cite{Note1}.

\begin{figure}
\includegraphics{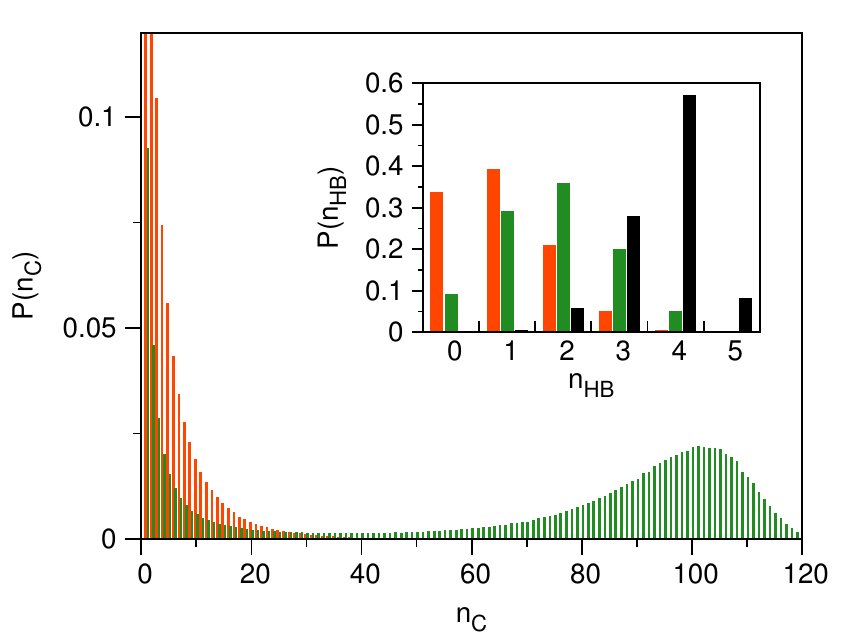}
\caption{\label{fig:hbs}
Probability that a water molecule belongs to a cluster of size $n_\mathrm{C}$
for LD--SCW (red, where $P(n_\mathrm{C} = 1)=34$\% and $P(2)=17$\%) and
HD--SCW (green).
The inset shows the distribution of the water molecules according to the
number of H-bonds formed including RTW (black).}
\end{figure}

In Fig.~\ref{fig:hbs} we analyze the clustering properties of SCW
at the two state points compared to RTW.
While for LD--SCW the relative abundance of the clusters decays almost
exponentially with the cluster size,
having highest values for water monomers followed by dimers,
the distribution for HD--SCW is clearly
bimodal with one maximum at non--H-bonded individual molecules and another one
close to the total available number of water molecules;
these $P(n_\mathrm{C})$ distributions are in good agreement with previous
simulations \cite{Churakov1999}.
In the inset of Fig.~\ref{fig:hbs} the propensity of the water molecules
to form H-bonds is illustrated.
For LD--SCW, non--H-bonded and singly H-bonded molecules predominate, while
this shifts to singly and doubly H-bonded ones at the higher density.
Importantly, the H-bond statistics in both cases is markedly different from
RTW, where four-coordinated molecules form a dense network
(i.e. a single ``cluster'' so that $P(n_\mathrm{C} = 128) \approx 1$)
and $\langle n_\mathrm{HB} \rangle = 3.68 \pm 0.07$.
For comparison:
$\langle n_\mathrm{HB} \rangle = 1.84 \pm 0.11$ and $1.00 \pm 0.12$
for HD-- and LD--SCW, respectively,
in agreement with other simulations \cite{Kalinichev1994, Ma2011}.
We checked on larger systems that these conclusions
are stable when increasing the system size~\cite{Note1}.
An independent experimental confirmation of the
dramatic reduction of H-bonding in SCW
comes from the Compton profiles obtained from IXS~\cite{Sit2007} that
provide indirect access to $\langle n_\mathrm{HB} \rangle$
via the number of electrons involved in hydrogen bonding.
The $\langle n_\mathrm{HB} \rangle$ value is then found to decrease
from 3.40 for RTW to 0.73 for SCW
(at $\rho=0.4$~g/cm$^{3}$ and 670~K)~\cite{Sit2007} in
agreement with our findings.

The significant changes in the fluid structure of SCW are expected to alter
significantly its IR~spectrum \cite{Tassaing2002,Kandratsenka2008}.
Particularly in the THz~range, being utmost sensitive to H-bond dynamics,
temperature increase is known to lead to major changes as observed
experimentally already at {\em subcritical} conditions.
Below the CP,
the two main features of the RTW spectrum, that is the librational mode at
$\approx 680$~cm$^{-1}$ ($\approx 20$~THz)
and the intermolecular H-bond stretching mode at $\approx 200$~cm$^{-1}$
(6~THz),
are known to merge gradually into a single feature centered at
$\approx 450$~cm$^{-1}$ ($\approx 13.5$~THz)
upon increasing the temperature to 510~K along
the gas--liquid coexistence curve \cite{Tassaing2002};
note that SCW spectra in the THz~range have not yet been measured.

\begin{figure}
\includegraphics{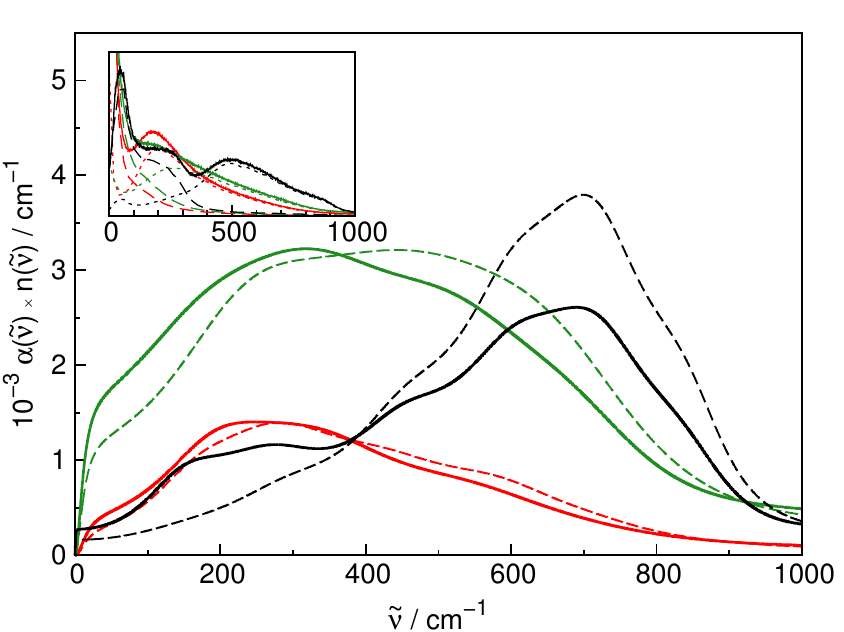}
\caption{\label{fig:ir}
Computed total THz~absorption spectra of water at the LD--SCW (red),
HD--SCW (green), and RTW (black) state points.
The Torii--corrected~\cite{Torii2014}  spectra (solid lines)
are compared to the uncorrected spectra (dashed lines), see text.
The inset shows the corresponding vibrational density of states
with separate contributions due to O~(dashed lines) and H~(dotted lines) atoms.
The zero-frequency peak of LD-- and HD--SCW is off scale
and the data are magnified in Fig.~S4 \cite{Note1}.}
\end{figure}

Our THz~spectra computed~\cite{Note1}
from the Fourier transform of the dipole derivative
time correlation function (TCF) are shown in Fig.~\ref{fig:ir}.
It is well known that flexible force fields without polarizability
and/or charge transfer cannot reproduce the THz~spectra of RTW,
in particular not the eminent H-bond network mode at $\approx 200$~cm$^{-1}$
(6~THz).
Several schemes to correct for this deficiency have been proposed and both
so-called ``Torii corrections'' \cite{Torii2011, Torii2014}
reach a considerable agreement with experimental and AIMD
IR~spectra of RTW~\cite{Heyden2010}.
In Torii's schemes,
which we generalized in the SM~\cite{Note1} to flexible water models,
polarization effects due to H-bonding are incorporated.
In the more advanced version~\cite{Torii2014}, this is achieved by adding
both intermolecular charge fluxes to existing H-bonds
and local electric field effects to each molecule when computing the
total dipole moment according to Eqs.~(S25)--(S28).
We scrutinized \cite{Note1} the applicability of these corrections
to SCW and conclude that they can be neglected in this regime,
which is backed up by comparison to available spectra
(see Sec.~IV.E and Fig.~S7 \cite{Note1}), in stark contrast to RTW.
The THz~spectra obtained with the most recent Torii correction \cite{Torii2014}
as well as the uncorrected original spectra clearly illustrate this point,
see Fig.~\ref{fig:ir}.

In the supercritical regime the THz~lineshape function
changes dramatically compared to ambient conditions:
it becomes unimodal in the first place and centered at
roughly 300~cm$^{-1}$ (9~THz)
at high density, while for LD--SCW it shifts down to about 200~cm$^{-1}$
(6~THz) according to Fig.~\ref{fig:ir}.
Even more pronounced are the changes in the vibrational density of states
(VDOS), see the inset or Fig.~S4~\cite{Note1}.
The librational band at $\approx 500$~cm$^{-1}$ (15~THz)
disappears for SCW in both the gas-like and liquid-like dynamical regimes,
while the principal peak due to the oxygen atoms movement at
$<100$~cm$^{-1}$ ($<3$~THz)
shifts to zero frequency and dramatically increases
in intensity, reflecting the only slightly hindered
{\em translational} (``ballistic'') movement
of water molecules in the supercritical fluid.
The latter is also evident from the value of the self-diffusion coefficient
estimated via the Green-Kubo relation from the center-of-mass velocity TCF.
It increases from 0.23 for RTW to 4.92 for HD--SCW
to 14.17~\r{A}$^{2}$/ps for LD--SCW
in accordance with the available QENS
data~\cite{Tassaing2000} (7.65 and 13.70~\r{A}$^{2}$/ps, respectively).
Considering separately the O/H atoms VDOS
(as defined in Sec.~III of the SM~\cite{Note1}),
the strikingly changing nature of the $\approx 200$~cm$^{-1}$
(6~THz) band is revealed.
While in RTW it is mainly due to oxygen motion, in SCW it is the hydrogen
motion that underlies this vibrational activity due to much more
{\em rotational} freedom of the individual water molecules.
This already gives a glimpse into the fundamentally different
nature of this
resonance at supercritical versus ambient conditions, vide infra.

The intra- and intermolecular correlations underlying both the IR~spectra
and VDOS can be resolved when applying a spatial
decomposition scheme \cite{Heyden2010}.
To this end, either molecular dipole velocities or mass-weighted atomic
velocities are projected on a grid to obtain spatially
resolved charge current or mass-weighted velocity density vector fields
which are then auto-correlated \cite{Note1}.
For isotropic liquids, such as RTW and SCW, angular averaging
leaves only a radial dependence of the spectrum,
at each frequency, on the distance from the reference point.
The resulting radially-resolved spectral response might be understood
most easily in analogy to RDFs:
while the latter describe radial correlations
\emph{in particle density} with increasing separation $r$,
the former captures \emph{vibrational correlations}
(in IR or particle dynamics) in a similar manner.

\begin{figure*}
\includegraphics{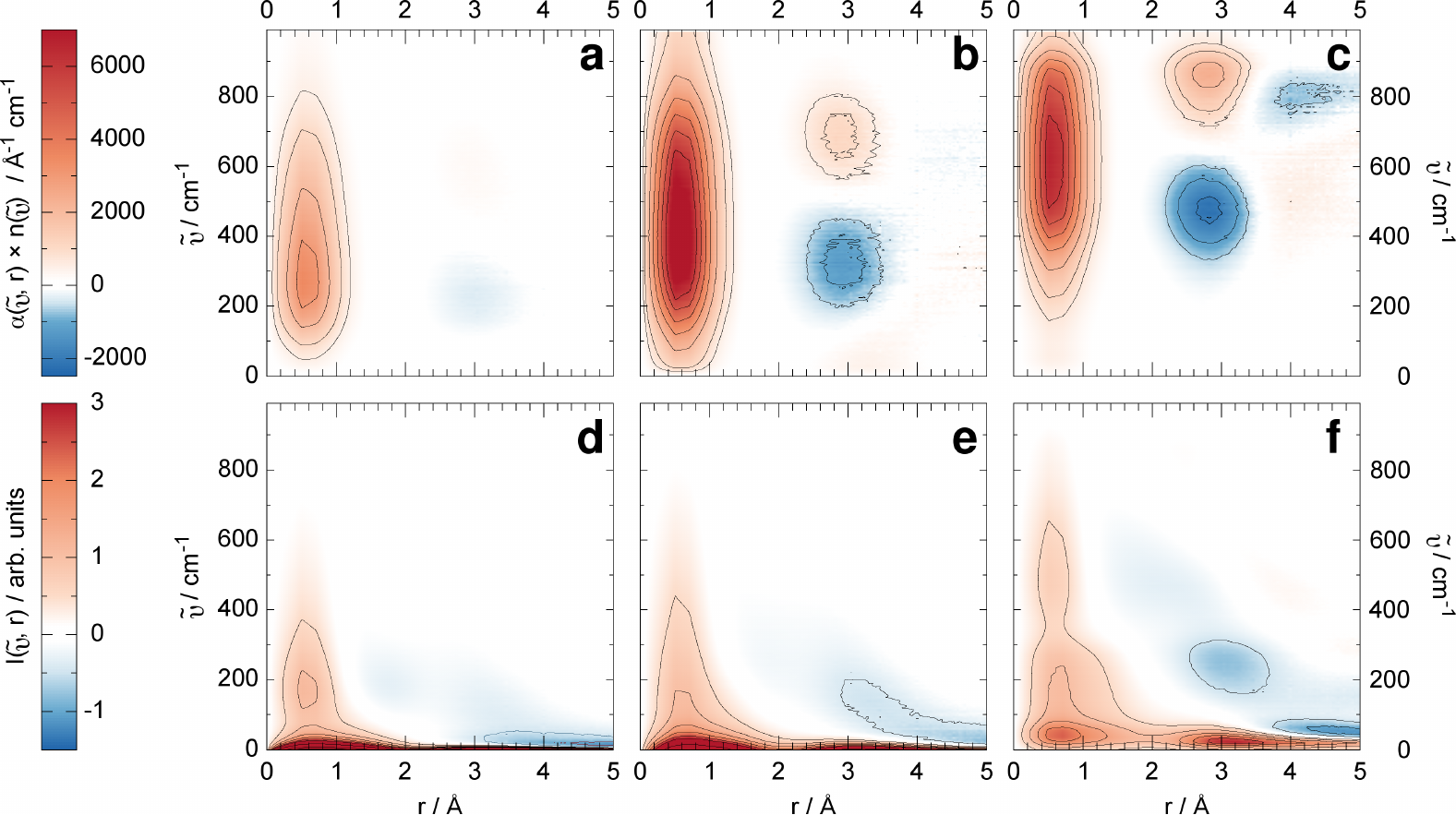}
\caption{\label{fig:radial}
Top: Radially-resolved THz~absorption spectra of water at the (a) LD--SCW,
(b) HD--SCW, and (c) RTW state points.
Bottom: Corresponding radially-resolved vibrational correlations
(``generalized VDOS'') at the (d) LD--SCW, (e) HD--SCW,
and (f) RTW state points.
Figure~S8 provides a three-dimensional representation of the
same data~\cite{Note1}.}
\end{figure*}

The radially-resolved IR~spectra of water reveal in the THz~spectral range
a characteristic loss of spatial correlations with decreasing density,
see panels \textbf{a}--\textbf{c} of Fig.~\ref{fig:radial}.
While for RTW, dipolar correlations up to the second hydration sphere
are clearly detectable in qualitative agreement with previous AIMD
analyses~\cite{Heyden2010},
for HD--SCW only the cross-correlations within the
first hydration sphere at roughly 3~\r{A} are important.
Even these short-range correlations fade away upon further
expansion of supercritical water to LD~conditions.
For LD--SCW, the major correlation features at this distance become less
intense and steadily red-shift relative to RTW by as much as 250~cm$^{-1}$.
Such red-shifts of the positive and negative {\em cross-correlation} peaks
reflect the increased librational freedom of the water molecules in the
supercritical state due to the lower degree of H-bonding
upon decreasing the density.
This general trend is in line with what has been observed in the
{\em subcritical} regime \cite{Tassaing2002} upon raising the temperature.

In the {\em supercritical} regime,
our findings therefore extend the Widom line crossover concept
from thermodynamics and dynamics \cite{Simeoni2010,McMillan2010,Gallo2014}
now into correlated dipolar response and THz~vibrational spectra.
The presence of weak correlations in the radially-resolved spectra of SCW
suggests that some nearest-neighbor dipolar couplings are
still effective even in supercritical water, while the long-range correlations
are strongly suppressed in the LD--SCW case due to
the completely disrupted tetrahedral H-bond network of water.
%
This directly relates the THz~response of
gas-like and liquid-like SCW to the percolation line~\cite{Bernabei2008b}.

Contrasting next the radially-resolved
vibrational correlations (see panels \textbf{d}--\textbf{f}
in Fig.~\ref{fig:radial}) to the dipolar THz~response
(panels \textbf{a}--\textbf{c})
it is striking to find out that the distinctly
modulated pattern characteristic for RTW
becomes greatly simplified at supercritical conditions.
With reference to RTW~\cite{Heyden2010}, this implies for SCW
the progressively diminishing spatial range and magnitude of cross-correlations
of the H-bond umbrella motion at ca. 80~cm$^{-1}$ (2.4~THz),
H-bond network stretching at 200~cm$^{-1}$ (6~THz),
librations at roughly 500~cm$^{-1}$ (15~THz),
as well as the absence of any noteworthy correlations
in the intramolecular vibrations range at still higher (mid--IR) frequencies.
This reflects the radical changes in the nature of the
underlying atomic motion at supercritical compared to ambient
conditions.
The steeply rising positive correlations close to
zero frequency can be attributed to increasingly ``ballistic''
movement of those water molecules which are no
longer firmly engaged in an extended H-bond network in SCW.
This pseudo-hard sphere character of water molecules in SCW and the long-range
nature of the observed velocity cross-correlations are reminiscent of the
similarly far-reaching correlations in the sedimentation of colloidal particles
due to instantaneous density fluctuations \cite{Segre1997}.
Next, the weak and broad feature at 200--300~cm$^{-1}$ does not
show much correlations even with nearest-neighbor molecules
(expected at $r \approx 3$~{\AA}), but instead it is
due to single-molecule motion (visible at $r \approx 0$~{\AA})
mainly of H~atoms (see Fig.~S4 in the SM~\cite{Note1}).
This librational single-particle motion is distinctly different from the
``network mode'' of RTW at 200~cm$^{-1}$ (6~THz)
that is caused by intermolecular stretching motion of tetrahedrally H-bonded
water molecules~\cite{Heyden2010} as evidenced by the prominent
cross-correlations at $r \approx 3$~{\AA}
(panel \textbf{f} in Fig.~\ref{fig:radial}).
Our peak assignment of the spectra in Figs.~4 and~3
is strongly supported by QENS experiments~\cite{Tassaing2000}
which provide a DOS for protons (see Fig.~9 therein)
that features a clearly structured response with two
pronounced maxima
at roughly 60 and 520~cm$^{-1}$ (7 and 65~meV) for RTW,
while only one broad unimodal resonance is observed for HD--SCW
(at $\rho=0.59$~g/cm$^{3}$ and 653~K)
close to 240~cm$^{-1}$ ($\approx 30$~meV).
This nicely agrees with our partial VDOS for H~atoms as depicted
in Fig.~S4~\cite{Note1}
with two peaks at about 45 and 500~cm$^{-1}$ for RTW and
a single broadband feature at 275~cm$^{-1}$ for HD--SCW.

In conclusion, dramatic changes of the THz~response of supercritical water
are predicted in this spectral region characteristic of the
tetrahedral H-bond network of liquid water, which are traced back to
significantly decreasing long-range intermolecular correlations.
In particular, the celebrated ``network mode'' of ambient water
at around 200~cm$^{-1}$ (6~THz)
vanishes at supercritical conditions, while the
libration band red-shifts by as much as 250~cm$^{-1}$.
Thus, instead of being bimodal, the supercritical THz~spectra
show a single but unusually broad, mainly librational feature
around 200--300~cm$^{-1}$.
Understanding the molecular mechanism that leads to the observed
THz~resonances will be key to interpret future experiments
of supercritical water and solutions.
In particular, the far-reaching couplings
characteristic to the THz~response of ambient liquid water
are found to be suppressed when
crossing the percolation line by moving from high to low densities
and thus from liquid-like to gas-like supercritical water, respectively.
The associated changes of the H-bond network motion
across the Widom line
are encoded in changes of the correlated dipolar response
as probed by THz~vibrational spectroscopy,
which therefore complements thermodynamic and dynamic techniques
to investigate this crossover concept in supercritical water
and beyond.
Finally, THz spectroscopy with its ability to detect changes in
the H-bond network between high-- and low-density water
might be an ideal technique to also
shed light on the Widom line in the framework of the controversially
discussed (second) liquid-liquid critical point of supercooled water.

\begin{acknowledgments}
We are particularly grateful to Hajime Torii
for clarifying the details of his IR~corrections.
We acknowledge partial financial support from DFG (MA~1547/11
and Cluster of Excellence ``RESOLV'' EXC~1069).
Computations were performed at LRZ M\"unchen, \textsc{HPC--RESOLV@RUB},
\textsc{Bovilab@RUB}, and RV--NRW.
\end{acknowledgments}


\begin{thebibliography}{58}%
\makeatletter
\providecommand \@ifxundefined [1]{%
 \@ifx{#1\undefined}
}%
\providecommand \@ifnum [1]{%
 \ifnum #1\expandafter \@firstoftwo
 \else \expandafter \@secondoftwo
 \fi
}%
\providecommand \@ifx [1]{%
 \ifx #1\expandafter \@firstoftwo
 \else \expandafter \@secondoftwo
 \fi
}%
\providecommand \natexlab [1]{#1}%
\providecommand \enquote  [1]{``#1''}%
\providecommand \bibnamefont  [1]{#1}%
\providecommand \bibfnamefont [1]{#1}%
\providecommand \citenamefont [1]{#1}%
\providecommand \href@noop [0]{\@secondoftwo}%
\providecommand \href [0]{\begingroup \@sanitize@url \@href}%
\providecommand \@href[1]{\@@startlink{#1}\@@href}%
\providecommand \@@href[1]{\endgroup#1\@@endlink}%
\providecommand \@sanitize@url [0]{\catcode `\\12\catcode `\$12\catcode
  `\&12\catcode `\#12\catcode `\^12\catcode `\_12\catcode `\%12\relax}%
\providecommand \@@startlink[1]{}%
\providecommand \@@endlink[0]{}%
\providecommand \url  [0]{\begingroup\@sanitize@url \@url }%
\providecommand \@url [1]{\endgroup\@href {#1}{\urlprefix }}%
\providecommand \urlprefix  [0]{URL }%
\providecommand \Eprint [0]{\href }%
\providecommand \doibase [0]{http://dx.doi.org/}%
\providecommand \selectlanguage [0]{\@gobble}%
\providecommand \bibinfo  [0]{\@secondoftwo}%
\providecommand \bibfield  [0]{\@secondoftwo}%
\providecommand \translation [1]{[#1]}%
\providecommand \BibitemOpen [0]{}%
\providecommand \bibitemStop [0]{}%
\providecommand \bibitemNoStop [0]{.\EOS\space}%
\providecommand \EOS [0]{\spacefactor3000\relax}%
\providecommand \BibitemShut  [1]{\csname bibitem#1\endcsname}%
\let\auto@bib@innerbib\@empty
\bibitem [{\citenamefont {Cunsolo}\ \emph {et~al.}(1998)\citenamefont
  {Cunsolo}, \citenamefont {Pratesi}, \citenamefont {Ruocco}, \citenamefont
  {Sampoli}, \citenamefont {Sette}, \citenamefont {Verbeni}, \citenamefont
  {Barocchi}, \citenamefont {Krisch}, \citenamefont {Masciovecchio},\ and\
  \citenamefont {Nardone}}]{Cunsolo1998}%
  \BibitemOpen
  \bibfield  {author} {\bibinfo {author} {\bibfnamefont {A.}~\bibnamefont
  {Cunsolo}}, \bibinfo {author} {\bibfnamefont {G.}~\bibnamefont {Pratesi}},
  \bibinfo {author} {\bibfnamefont {G.}~\bibnamefont {Ruocco}}, \bibinfo
  {author} {\bibfnamefont {M.}~\bibnamefont {Sampoli}}, \bibinfo {author}
  {\bibfnamefont {F.}~\bibnamefont {Sette}}, \bibinfo {author} {\bibfnamefont
  {R.}~\bibnamefont {Verbeni}}, \bibinfo {author} {\bibfnamefont
  {F.}~\bibnamefont {Barocchi}}, \bibinfo {author} {\bibfnamefont
  {M.}~\bibnamefont {Krisch}}, \bibinfo {author} {\bibfnamefont
  {C.}~\bibnamefont {Masciovecchio}}, \ and\ \bibinfo {author} {\bibfnamefont
  {M.}~\bibnamefont {Nardone}},\ }\href@noop {} {\bibfield  {journal} {\bibinfo
   {journal} {Phys. Rev. Lett.}\ }\textbf {\bibinfo {volume} {80}},\ \bibinfo
  {pages} {3515} (\bibinfo {year} {1998})}\BibitemShut {NoStop}%
\bibitem [{\citenamefont {Gorelli}\ \emph {et~al.}(2006)\citenamefont
  {Gorelli}, \citenamefont {Santoro}, \citenamefont {Scopigno}, \citenamefont
  {Krisch},\ and\ \citenamefont {Ruocco}}]{Gorelli2006}%
  \BibitemOpen
  \bibfield  {author} {\bibinfo {author} {\bibfnamefont {F.}~\bibnamefont
  {Gorelli}}, \bibinfo {author} {\bibfnamefont {M.}~\bibnamefont {Santoro}},
  \bibinfo {author} {\bibfnamefont {T.}~\bibnamefont {Scopigno}}, \bibinfo
  {author} {\bibfnamefont {M.}~\bibnamefont {Krisch}}, \ and\ \bibinfo {author}
  {\bibfnamefont {G.}~\bibnamefont {Ruocco}},\ }\href@noop {} {\bibfield
  {journal} {\bibinfo  {journal} {Phys. Rev. Lett.}\ }\textbf {\bibinfo
  {volume} {97}},\ \bibinfo {pages} {245702} (\bibinfo {year}
  {2006})}\BibitemShut {NoStop}%
\bibitem [{\citenamefont {Simeoni}\ \emph {et~al.}(2010)\citenamefont
  {Simeoni}, \citenamefont {Bryk}, \citenamefont {Gorelli}, \citenamefont
  {Krisch}, \citenamefont {Ruocco}, \citenamefont {Santoro},\ and\
  \citenamefont {Scopigno}}]{Simeoni2010}%
  \BibitemOpen
  \bibfield  {author} {\bibinfo {author} {\bibfnamefont {G.~G.}\ \bibnamefont
  {Simeoni}}, \bibinfo {author} {\bibfnamefont {T.}~\bibnamefont {Bryk}},
  \bibinfo {author} {\bibfnamefont {F.~A.}\ \bibnamefont {Gorelli}}, \bibinfo
  {author} {\bibfnamefont {M.}~\bibnamefont {Krisch}}, \bibinfo {author}
  {\bibfnamefont {G.}~\bibnamefont {Ruocco}}, \bibinfo {author} {\bibfnamefont
  {M.}~\bibnamefont {Santoro}}, \ and\ \bibinfo {author} {\bibfnamefont
  {T.}~\bibnamefont {Scopigno}},\ }\href@noop {} {\bibfield  {journal}
  {\bibinfo  {journal} {Nat. Phys.}\ }\textbf {\bibinfo {volume} {6}},\
  \bibinfo {pages} {503} (\bibinfo {year} {2010})}\BibitemShut {NoStop}%
\bibitem [{\citenamefont {McMillan}\ and\ \citenamefont
  {Stanley}(2010)}]{McMillan2010}%
  \BibitemOpen
  \bibfield  {author} {\bibinfo {author} {\bibfnamefont {P.~F.}\ \bibnamefont
  {McMillan}}\ and\ \bibinfo {author} {\bibfnamefont {H.~E.}\ \bibnamefont
  {Stanley}},\ }\href@noop {} {\bibfield  {journal} {\bibinfo  {journal} {Nat.
  Phys.}\ }\textbf {\bibinfo {volume} {6}},\ \bibinfo {pages} {479} (\bibinfo
  {year} {2010})}\BibitemShut {NoStop}%
\bibitem [{\citenamefont {Bolmatov}\ \emph {et~al.}(2013)\citenamefont
  {Bolmatov}, \citenamefont {Brazhkin},\ and\ \citenamefont
  {Trachenko}}]{Bolmatov2013}%
  \BibitemOpen
  \bibfield  {author} {\bibinfo {author} {\bibfnamefont {D.}~\bibnamefont
  {Bolmatov}}, \bibinfo {author} {\bibfnamefont {V.}~\bibnamefont {Brazhkin}},
  \ and\ \bibinfo {author} {\bibfnamefont {K.}~\bibnamefont {Trachenko}},\
  }\href@noop {} {\bibfield  {journal} {\bibinfo  {journal} {Nat. Commun.}\
  }\textbf {\bibinfo {volume} {4}},\ \bibinfo {pages} {2331} (\bibinfo {year}
  {2013})}\BibitemShut {NoStop}%
\bibitem [{\citenamefont {Gallo}\ \emph {et~al.}(2014)\citenamefont {Gallo},
  \citenamefont {Corradini},\ and\ \citenamefont {Rovere}}]{Gallo2014}%
  \BibitemOpen
  \bibfield  {author} {\bibinfo {author} {\bibfnamefont {P.}~\bibnamefont
  {Gallo}}, \bibinfo {author} {\bibfnamefont {D.}~\bibnamefont {Corradini}}, \
  and\ \bibinfo {author} {\bibfnamefont {M.}~\bibnamefont {Rovere}},\
  }\href@noop {} {\bibfield  {journal} {\bibinfo  {journal} {Nat. Commun.}\
  }\textbf {\bibinfo {volume} {5}},\ \bibinfo {pages} {5806} (\bibinfo {year}
  {2014})}\BibitemShut {NoStop}%
\bibitem [{\citenamefont {Sokhan}\ \emph {et~al.}(2015)\citenamefont {Sokhan},
  \citenamefont {Jones}, \citenamefont {Cipcigan}, \citenamefont {Crain},\ and\
  \citenamefont {Martyna}}]{Sokhan2015}%
  \BibitemOpen
  \bibfield  {author} {\bibinfo {author} {\bibfnamefont {V.~P.}\ \bibnamefont
  {Sokhan}}, \bibinfo {author} {\bibfnamefont {A.}~\bibnamefont {Jones}},
  \bibinfo {author} {\bibfnamefont {F.~S.}\ \bibnamefont {Cipcigan}}, \bibinfo
  {author} {\bibfnamefont {J.}~\bibnamefont {Crain}}, \ and\ \bibinfo {author}
  {\bibfnamefont {G.~J.}\ \bibnamefont {Martyna}},\ }\href@noop {} {\bibfield
  {journal} {\bibinfo  {journal} {Phys. Rev. Lett.}\ }\textbf {\bibinfo
  {volume} {115}},\ \bibinfo {pages} {117801} (\bibinfo {year}
  {2015})}\BibitemShut {NoStop}%
\bibitem [{\citenamefont {Fomin}\ \emph {et~al.}(2015)\citenamefont {Fomin},
  \citenamefont {Ryzhov}, \citenamefont {Tsiok},\ and\ \citenamefont
  {Brazhkin}}]{Fomin2015}%
  \BibitemOpen
  \bibfield  {author} {\bibinfo {author} {\bibfnamefont {Y.~D.}\ \bibnamefont
  {Fomin}}, \bibinfo {author} {\bibfnamefont {V.~N.}\ \bibnamefont {Ryzhov}},
  \bibinfo {author} {\bibfnamefont {E.~N.}\ \bibnamefont {Tsiok}}, \ and\
  \bibinfo {author} {\bibfnamefont {V.~V.}\ \bibnamefont {Brazhkin}},\
  }\href@noop {} {\bibfield  {journal} {\bibinfo  {journal} {Sci. Rep.}\
  }\textbf {\bibinfo {volume} {5}},\ \bibinfo {pages} {14234} (\bibinfo {year}
  {2015})}\BibitemShut {NoStop}%
\bibitem [{\citenamefont {Li}\ and\ \citenamefont {Trost}(2008)}]{Li2008}%
  \BibitemOpen
  \bibfield  {author} {\bibinfo {author} {\bibfnamefont {C.-J.}\ \bibnamefont
  {Li}}\ and\ \bibinfo {author} {\bibfnamefont {B.~M.}\ \bibnamefont {Trost}},\
  }\href@noop {} {\bibfield  {journal} {\bibinfo  {journal} {Proc. Natl. Acad.
  Sci. U. S. A.}\ }\textbf {\bibinfo {volume} {105}},\ \bibinfo {pages} {13197}
  (\bibinfo {year} {2008})}\BibitemShut {NoStop}%
\bibitem [{\citenamefont {DeSimone}(2002)}]{DeSimone2002}%
  \BibitemOpen
  \bibfield  {author} {\bibinfo {author} {\bibfnamefont {J.~M.}\ \bibnamefont
  {DeSimone}},\ }\href@noop {} {\bibfield  {journal} {\bibinfo  {journal}
  {Science}\ }\textbf {\bibinfo {volume} {297}},\ \bibinfo {pages} {799}
  (\bibinfo {year} {2002})}\BibitemShut {NoStop}%
\bibitem [{\citenamefont {Jessop}\ \emph {et~al.}(1995)\citenamefont {Jessop},
  \citenamefont {Ikariya},\ and\ \citenamefont {Noyori}}]{Jessop1995}%
  \BibitemOpen
  \bibfield  {author} {\bibinfo {author} {\bibfnamefont {P.~G.}\ \bibnamefont
  {Jessop}}, \bibinfo {author} {\bibfnamefont {T.}~\bibnamefont {Ikariya}}, \
  and\ \bibinfo {author} {\bibfnamefont {R.}~\bibnamefont {Noyori}},\
  }\href@noop {} {\bibfield  {journal} {\bibinfo  {journal} {Science}\ }\textbf
  {\bibinfo {volume} {269}},\ \bibinfo {pages} {1065} (\bibinfo {year}
  {1995})}\BibitemShut {NoStop}%
\bibitem [{\citenamefont {Pan}\ \emph {et~al.}(2013)\citenamefont {Pan},
  \citenamefont {Spanu}, \citenamefont {Harrison}, \citenamefont {Sverjensky},\
  and\ \citenamefont {Galli}}]{Pan2013}%
  \BibitemOpen
  \bibfield  {author} {\bibinfo {author} {\bibfnamefont {D.}~\bibnamefont
  {Pan}}, \bibinfo {author} {\bibfnamefont {L.}~\bibnamefont {Spanu}}, \bibinfo
  {author} {\bibfnamefont {B.}~\bibnamefont {Harrison}}, \bibinfo {author}
  {\bibfnamefont {D.~A.}\ \bibnamefont {Sverjensky}}, \ and\ \bibinfo {author}
  {\bibfnamefont {G.}~\bibnamefont {Galli}},\ }\href@noop {} {\bibfield
  {journal} {\bibinfo  {journal} {Proc. Natl. Acad. Sci. U. S. A.}\ }\textbf
  {\bibinfo {volume} {110}},\ \bibinfo {pages} {6646} (\bibinfo {year}
  {2013})}\BibitemShut {NoStop}%
\bibitem [{\citenamefont {Cavazzoni}\ \emph {et~al.}(1999)\citenamefont
  {Cavazzoni}, \citenamefont {Chiarotti}, \citenamefont {Scandolo},
  \citenamefont {Tosatti}, \citenamefont {Bernasconi},\ and\ \citenamefont
  {Parrinello}}]{Cavazzoni1999}%
  \BibitemOpen
  \bibfield  {author} {\bibinfo {author} {\bibfnamefont {C.}~\bibnamefont
  {Cavazzoni}}, \bibinfo {author} {\bibfnamefont {G.~L.}\ \bibnamefont
  {Chiarotti}}, \bibinfo {author} {\bibfnamefont {S.}~\bibnamefont {Scandolo}},
  \bibinfo {author} {\bibfnamefont {E.}~\bibnamefont {Tosatti}}, \bibinfo
  {author} {\bibfnamefont {M.}~\bibnamefont {Bernasconi}}, \ and\ \bibinfo
  {author} {\bibfnamefont {M.}~\bibnamefont {Parrinello}},\ }\href@noop {}
  {\bibfield  {journal} {\bibinfo  {journal} {Science}\ }\textbf {\bibinfo
  {volume} {283}},\ \bibinfo {pages} {44} (\bibinfo {year} {1999})}\BibitemShut
  {NoStop}%
\bibitem [{\citenamefont {{Weing\"artner}}\ and\ \citenamefont
  {Franck}(2005)}]{Weingartner2005}%
  \BibitemOpen
  \bibfield  {author} {\bibinfo {author} {\bibfnamefont {H.}~\bibnamefont
  {{Weing\"artner}}}\ and\ \bibinfo {author} {\bibfnamefont {E.~U.}\
  \bibnamefont {Franck}},\ }\href@noop {} {\bibfield  {journal} {\bibinfo
  {journal} {Angew. Chem. Int. Ed.}\ }\textbf {\bibinfo {volume} {44}},\
  \bibinfo {pages} {2672} (\bibinfo {year} {2005})}\BibitemShut {NoStop}%
\bibitem [{\citenamefont {Postorino}\ \emph {et~al.}(1993)\citenamefont
  {Postorino}, \citenamefont {Tromp}, \citenamefont {Ricci}, \citenamefont
  {Soper},\ and\ \citenamefont {Neilson}}]{Postorino1993}%
  \BibitemOpen
  \bibfield  {author} {\bibinfo {author} {\bibfnamefont {P.}~\bibnamefont
  {Postorino}}, \bibinfo {author} {\bibfnamefont {R.~H.}\ \bibnamefont
  {Tromp}}, \bibinfo {author} {\bibfnamefont {M.-A.}\ \bibnamefont {Ricci}},
  \bibinfo {author} {\bibfnamefont {A.~K.}\ \bibnamefont {Soper}}, \ and\
  \bibinfo {author} {\bibfnamefont {G.~W.}\ \bibnamefont {Neilson}},\
  }\href@noop {} {\bibfield  {journal} {\bibinfo  {journal} {Nature}\ }\textbf
  {\bibinfo {volume} {366}},\ \bibinfo {pages} {668} (\bibinfo {year}
  {1993})}\BibitemShut {NoStop}%
\bibitem [{\citenamefont {Ohtaki}\ \emph {et~al.}(1997)\citenamefont {Ohtaki},
  \citenamefont {Radnai},\ and\ \citenamefont {Yamaguchi}}]{Ohtaki1997}%
  \BibitemOpen
  \bibfield  {author} {\bibinfo {author} {\bibfnamefont {H.}~\bibnamefont
  {Ohtaki}}, \bibinfo {author} {\bibfnamefont {T.}~\bibnamefont {Radnai}}, \
  and\ \bibinfo {author} {\bibfnamefont {T.}~\bibnamefont {Yamaguchi}},\
  }\href@noop {} {\bibfield  {journal} {\bibinfo  {journal} {Chem. Soc. Rev.}\
  }\textbf {\bibinfo {volume} {26}},\ \bibinfo {pages} {41} (\bibinfo {year}
  {1997})}\BibitemShut {NoStop}%
\bibitem [{\citenamefont {Bernabei}\ \emph {et~al.}(2008)\citenamefont
  {Bernabei}, \citenamefont {Botti}, \citenamefont {Bruni}, \citenamefont
  {Ricci},\ and\ \citenamefont {Soper}}]{Bernabei2008b}%
  \BibitemOpen
  \bibfield  {author} {\bibinfo {author} {\bibfnamefont {M.}~\bibnamefont
  {Bernabei}}, \bibinfo {author} {\bibfnamefont {A.}~\bibnamefont {Botti}},
  \bibinfo {author} {\bibfnamefont {F.}~\bibnamefont {Bruni}}, \bibinfo
  {author} {\bibfnamefont {M.-A.}\ \bibnamefont {Ricci}}, \ and\ \bibinfo
  {author} {\bibfnamefont {A.~K.}\ \bibnamefont {Soper}},\ }\href@noop {}
  {\bibfield  {journal} {\bibinfo  {journal} {Phys. Rev. E}\ }\textbf {\bibinfo
  {volume} {78}},\ \bibinfo {pages} {021505} (\bibinfo {year}
  {2008})}\BibitemShut {NoStop}%
\bibitem [{\citenamefont {Tassaing}\ and\ \citenamefont
  {Bellissent-Funel}(2000)}]{Tassaing2000}%
  \BibitemOpen
  \bibfield  {author} {\bibinfo {author} {\bibfnamefont {T.}~\bibnamefont
  {Tassaing}}\ and\ \bibinfo {author} {\bibfnamefont {M.-C.}\ \bibnamefont
  {Bellissent-Funel}},\ }\href@noop {} {\bibfield  {journal} {\bibinfo
  {journal} {J. Chem. Phys.}\ }\textbf {\bibinfo {volume} {113}},\ \bibinfo
  {pages} {3332} (\bibinfo {year} {2000})}\BibitemShut {NoStop}%
\bibitem [{\citenamefont {Sit}\ \emph {et~al.}(2007)\citenamefont {Sit},
  \citenamefont {Bellin}, \citenamefont {Barbiellini}, \citenamefont
  {Testemale}, \citenamefont {Hazemann}, \citenamefont {Buslaps}, \citenamefont
  {Marzari},\ and\ \citenamefont {Shukla}}]{Sit2007}%
  \BibitemOpen
  \bibfield  {author} {\bibinfo {author} {\bibfnamefont {P.~H.-L.}\
  \bibnamefont {Sit}}, \bibinfo {author} {\bibfnamefont {C.}~\bibnamefont
  {Bellin}}, \bibinfo {author} {\bibfnamefont {B.}~\bibnamefont {Barbiellini}},
  \bibinfo {author} {\bibfnamefont {D.}~\bibnamefont {Testemale}}, \bibinfo
  {author} {\bibfnamefont {J.-L.}\ \bibnamefont {Hazemann}}, \bibinfo {author}
  {\bibfnamefont {T.}~\bibnamefont {Buslaps}}, \bibinfo {author} {\bibfnamefont
  {N.}~\bibnamefont {Marzari}}, \ and\ \bibinfo {author} {\bibfnamefont
  {A.}~\bibnamefont {Shukla}},\ }\href@noop {} {\bibfield  {journal} {\bibinfo
  {journal} {Phys. Rev. B}\ }\textbf {\bibinfo {volume} {76}},\ \bibinfo
  {pages} {245413} (\bibinfo {year} {2007})}\BibitemShut {NoStop}%
\bibitem [{\citenamefont {Pantalei}\ \emph {et~al.}(2008)\citenamefont
  {Pantalei}, \citenamefont {Pietropaolo}, \citenamefont {Senesi},
  \citenamefont {Imberti}, \citenamefont {Andreani}, \citenamefont {Mayers},
  \citenamefont {Burnham},\ and\ \citenamefont {Reiter}}]{Pantalei2008}%
  \BibitemOpen
  \bibfield  {author} {\bibinfo {author} {\bibfnamefont {C.}~\bibnamefont
  {Pantalei}}, \bibinfo {author} {\bibfnamefont {A.}~\bibnamefont
  {Pietropaolo}}, \bibinfo {author} {\bibfnamefont {R.}~\bibnamefont {Senesi}},
  \bibinfo {author} {\bibfnamefont {S.}~\bibnamefont {Imberti}}, \bibinfo
  {author} {\bibfnamefont {C.}~\bibnamefont {Andreani}}, \bibinfo {author}
  {\bibfnamefont {J.}~\bibnamefont {Mayers}}, \bibinfo {author} {\bibfnamefont
  {C.}~\bibnamefont {Burnham}}, \ and\ \bibinfo {author} {\bibfnamefont
  {G.}~\bibnamefont {Reiter}},\ }\href@noop {} {\bibfield  {journal} {\bibinfo
  {journal} {Phys. Rev. Lett.}\ }\textbf {\bibinfo {volume} {100}},\ \bibinfo
  {pages} {177801} (\bibinfo {year} {2008})}\BibitemShut {NoStop}%
\bibitem [{\citenamefont {Sahle}\ \emph {et~al.}(2013)\citenamefont {Sahle},
  \citenamefont {Sternemann}, \citenamefont {Schmidt}, \citenamefont {Lehtola},
  \citenamefont {Jahn}, \citenamefont {Simonelli}, \citenamefont {Huotari},
  \citenamefont {Hakala}, \citenamefont {Pylkkanen}, \citenamefont {Nyrow},
  \citenamefont {Mende}, \citenamefont {Tolan}, \citenamefont {Hamalainen},\
  and\ \citenamefont {Wilke}}]{Sahle2013}%
  \BibitemOpen
  \bibfield  {author} {\bibinfo {author} {\bibfnamefont {C.~J.}\ \bibnamefont
  {Sahle}}, \bibinfo {author} {\bibfnamefont {C.}~\bibnamefont {Sternemann}},
  \bibinfo {author} {\bibfnamefont {C.}~\bibnamefont {Schmidt}}, \bibinfo
  {author} {\bibfnamefont {S.}~\bibnamefont {Lehtola}}, \bibinfo {author}
  {\bibfnamefont {S.}~\bibnamefont {Jahn}}, \bibinfo {author} {\bibfnamefont
  {L.}~\bibnamefont {Simonelli}}, \bibinfo {author} {\bibfnamefont
  {S.}~\bibnamefont {Huotari}}, \bibinfo {author} {\bibfnamefont
  {M.}~\bibnamefont {Hakala}}, \bibinfo {author} {\bibfnamefont
  {T.}~\bibnamefont {Pylkkanen}}, \bibinfo {author} {\bibfnamefont
  {A.}~\bibnamefont {Nyrow}}, \bibinfo {author} {\bibfnamefont
  {K.}~\bibnamefont {Mende}}, \bibinfo {author} {\bibfnamefont
  {M.}~\bibnamefont {Tolan}}, \bibinfo {author} {\bibfnamefont
  {K.}~\bibnamefont {Hamalainen}}, \ and\ \bibinfo {author} {\bibfnamefont
  {M.}~\bibnamefont {Wilke}},\ }\href@noop {} {\bibfield  {journal} {\bibinfo
  {journal} {Proc. Natl. Acad. Sci. U. S. A.}\ }\textbf {\bibinfo {volume}
  {110}},\ \bibinfo {pages} {6301} (\bibinfo {year} {2013})}\BibitemShut
  {NoStop}%
\bibitem [{\citenamefont {Matubayasi}\ \emph {et~al.}(1997)\citenamefont
  {Matubayasi}, \citenamefont {Wakai},\ and\ \citenamefont
  {Nakahara}}]{Matubayasi1997a}%
  \BibitemOpen
  \bibfield  {author} {\bibinfo {author} {\bibfnamefont {N.}~\bibnamefont
  {Matubayasi}}, \bibinfo {author} {\bibfnamefont {C.}~\bibnamefont {Wakai}}, \
  and\ \bibinfo {author} {\bibfnamefont {M.}~\bibnamefont {Nakahara}},\
  }\href@noop {} {\bibfield  {journal} {\bibinfo  {journal} {Phys. Rev. Lett.}\
  }\textbf {\bibinfo {volume} {78}},\ \bibinfo {pages} {2573} (\bibinfo {year}
  {1997})}\BibitemShut {NoStop}%
\bibitem [{\citenamefont {Kalinichev}(2001)}]{Kalinichev2001}%
  \BibitemOpen
  \bibfield  {author} {\bibinfo {author} {\bibfnamefont {A.~G.}\ \bibnamefont
  {Kalinichev}},\ }\href@noop {} {\bibfield  {journal} {\bibinfo  {journal}
  {Rev. Mineral. Geochem.}\ }\textbf {\bibinfo {volume} {42}},\ \bibinfo
  {pages} {83} (\bibinfo {year} {2001})}\BibitemShut {NoStop}%
\bibitem [{\citenamefont {Marx}\ and\ \citenamefont {Hutter}(2009)}]{Marx2009}%
  \BibitemOpen
  \bibfield  {author} {\bibinfo {author} {\bibfnamefont {D.}~\bibnamefont
  {Marx}}\ and\ \bibinfo {author} {\bibfnamefont {J.}~\bibnamefont {Hutter}},\
  }\href@noop {} {\emph {\bibinfo {title} {Ab Initio Molecular Dynamics}}}\
  (\bibinfo  {publisher} {Cambridge University Press},\ \bibinfo {address}
  {Cambridge},\ \bibinfo {year} {2009})\BibitemShut {NoStop}%
\bibitem [{\citenamefont {Boero}\ \emph {et~al.}(2000)\citenamefont {Boero},
  \citenamefont {Terakura}, \citenamefont {Ikeshoji}, \citenamefont {Liew},\
  and\ \citenamefont {Parrinello}}]{Boero2000}%
  \BibitemOpen
  \bibfield  {author} {\bibinfo {author} {\bibfnamefont {M.}~\bibnamefont
  {Boero}}, \bibinfo {author} {\bibfnamefont {K.}~\bibnamefont {Terakura}},
  \bibinfo {author} {\bibfnamefont {T.}~\bibnamefont {Ikeshoji}}, \bibinfo
  {author} {\bibfnamefont {C.~C.}\ \bibnamefont {Liew}}, \ and\ \bibinfo
  {author} {\bibfnamefont {M.}~\bibnamefont {Parrinello}},\ }\href@noop {}
  {\bibfield  {journal} {\bibinfo  {journal} {Phys. Rev. Lett.}\ }\textbf
  {\bibinfo {volume} {85}},\ \bibinfo {pages} {3245} (\bibinfo {year}
  {2000})}\BibitemShut {NoStop}%
\bibitem [{\citenamefont {Boero}\ \emph {et~al.}(2003)\citenamefont {Boero},
  \citenamefont {Parrinello}, \citenamefont {Terakura}, \citenamefont
  {Ikeshoji},\ and\ \citenamefont {Liew}}]{Boero2003}%
  \BibitemOpen
  \bibfield  {author} {\bibinfo {author} {\bibfnamefont {M.}~\bibnamefont
  {Boero}}, \bibinfo {author} {\bibfnamefont {M.}~\bibnamefont {Parrinello}},
  \bibinfo {author} {\bibfnamefont {K.}~\bibnamefont {Terakura}}, \bibinfo
  {author} {\bibfnamefont {T.}~\bibnamefont {Ikeshoji}}, \ and\ \bibinfo
  {author} {\bibfnamefont {C.~C.}\ \bibnamefont {Liew}},\ }\href@noop {}
  {\bibfield  {journal} {\bibinfo  {journal} {Phys. Rev. Lett.}\ }\textbf
  {\bibinfo {volume} {90}},\ \bibinfo {pages} {226403} (\bibinfo {year}
  {2003})}\BibitemShut {NoStop}%
\bibitem [{\citenamefont {Yagasaki}\ and\ \citenamefont
  {Saito}(2013)}]{Yagasaki2013}%
  \BibitemOpen
  \bibfield  {author} {\bibinfo {author} {\bibfnamefont {T.}~\bibnamefont
  {Yagasaki}}\ and\ \bibinfo {author} {\bibfnamefont {S.}~\bibnamefont
  {Saito}},\ }\href@noop {} {\bibfield  {journal} {\bibinfo  {journal} {Annu.
  Rev. Phys. Chem.}\ }\textbf {\bibinfo {volume} {64}},\ \bibinfo {pages} {55}
  (\bibinfo {year} {2013})}\BibitemShut {NoStop}%
\bibitem [{\citenamefont {Tassaing}\ \emph {et~al.}(2002)\citenamefont
  {Tassaing}, \citenamefont {Danten},\ and\ \citenamefont
  {Besnard}}]{Tassaing2002}%
  \BibitemOpen
  \bibfield  {author} {\bibinfo {author} {\bibfnamefont {T.}~\bibnamefont
  {Tassaing}}, \bibinfo {author} {\bibfnamefont {Y.}~\bibnamefont {Danten}}, \
  and\ \bibinfo {author} {\bibfnamefont {M.}~\bibnamefont {Besnard}},\
  }\href@noop {} {\bibfield  {journal} {\bibinfo  {journal} {J. Mol. Liq.}\
  }\textbf {\bibinfo {volume} {101}},\ \bibinfo {pages} {149} (\bibinfo {year}
  {2002})}\BibitemShut {NoStop}%
\bibitem [{\citenamefont {Kandratsenka}\ \emph {et~al.}(2008)\citenamefont
  {Kandratsenka}, \citenamefont {Schwarzer},\ and\ \citenamefont
  {{V\"ohringer}}}]{Kandratsenka2008}%
  \BibitemOpen
  \bibfield  {author} {\bibinfo {author} {\bibfnamefont {A.}~\bibnamefont
  {Kandratsenka}}, \bibinfo {author} {\bibfnamefont {D.}~\bibnamefont
  {Schwarzer}}, \ and\ \bibinfo {author} {\bibfnamefont {P.}~\bibnamefont
  {{V\"ohringer}}},\ }\href@noop {} {\bibfield  {journal} {\bibinfo  {journal}
  {J. Chem. Phys.}\ }\textbf {\bibinfo {volume} {128}},\ \bibinfo {pages}
  {244510} (\bibinfo {year} {2008})}\BibitemShut {NoStop}%
\bibitem [{\citenamefont {Chalmers}\ and\ \citenamefont
  {Griffiths}(2006)}]{Handbook2006}%
  \BibitemOpen
  \bibinfo {editor} {\bibfnamefont {J.~M.}\ \bibnamefont {Chalmers}}\ and\
  \bibinfo {editor} {\bibfnamefont {P.~R.}\ \bibnamefont {Griffiths}},\ eds.,\
  \href@noop {} {\emph {\bibinfo {title} {Handbook of Vibrational
  Spectroscopy}}},\ Vol.\ \bibinfo {volume} {1.~Theory and Instrumentation}\
  (\bibinfo  {publisher} {John Wiley {\&} Sons, Ltd},\ \bibinfo {year}
  {2006})\BibitemShut {NoStop}%
\bibitem [{\citenamefont {R{\o}nne}\ \emph {et~al.}(1999)\citenamefont
  {R{\o}nne}, \citenamefont {{\AA}strand},\ and\ \citenamefont
  {Keiding}}]{Ronne1999}%
  \BibitemOpen
  \bibfield  {author} {\bibinfo {author} {\bibfnamefont {C.}~\bibnamefont
  {R{\o}nne}}, \bibinfo {author} {\bibfnamefont {P.-O.}\ \bibnamefont
  {{\AA}strand}}, \ and\ \bibinfo {author} {\bibfnamefont {R.~S.}\ \bibnamefont
  {Keiding}},\ }\href@noop {} {\bibfield  {journal} {\bibinfo  {journal} {Phys.
  Rev. Lett.}\ }\textbf {\bibinfo {volume} {82}},\ \bibinfo {pages} {2888}
  (\bibinfo {year} {1999})}\BibitemShut {NoStop}%
\bibitem [{\citenamefont {Buck}\ and\ \citenamefont
  {Huisken}(2000)}]{Buck2000}%
  \BibitemOpen
  \bibfield  {author} {\bibinfo {author} {\bibfnamefont {U.}~\bibnamefont
  {Buck}}\ and\ \bibinfo {author} {\bibfnamefont {F.}~\bibnamefont {Huisken}},\
  }\href@noop {} {\bibfield  {journal} {\bibinfo  {journal} {Chem. Rev.}\
  }\textbf {\bibinfo {volume} {100}},\ \bibinfo {pages} {3863} (\bibinfo {year}
  {2000})}\BibitemShut {NoStop}%
\bibitem [{\citenamefont {Heugen}\ \emph {et~al.}(2006)\citenamefont {Heugen},
  \citenamefont {Schwaab}, \citenamefont {Br{\"u}ndermann}, \citenamefont
  {Heyden}, \citenamefont {Yu}, \citenamefont {Leitner},\ and\ \citenamefont
  {Havenith}}]{Heugen06}%
  \BibitemOpen
  \bibfield  {author} {\bibinfo {author} {\bibfnamefont {U.}~\bibnamefont
  {Heugen}}, \bibinfo {author} {\bibfnamefont {G.}~\bibnamefont {Schwaab}},
  \bibinfo {author} {\bibfnamefont {E.}~\bibnamefont {Br{\"u}ndermann}},
  \bibinfo {author} {\bibfnamefont {M.}~\bibnamefont {Heyden}}, \bibinfo
  {author} {\bibfnamefont {X.}~\bibnamefont {Yu}}, \bibinfo {author}
  {\bibfnamefont {D.-M.}\ \bibnamefont {Leitner}}, \ and\ \bibinfo {author}
  {\bibfnamefont {M.}~\bibnamefont {Havenith}},\ }\href@noop {} {\bibfield
  {journal} {\bibinfo  {journal} {Proc. Natl. Acad. Sci. U.S.A.}\ }\textbf
  {\bibinfo {volume} {103}},\ \bibinfo {pages} {12301} (\bibinfo {year}
  {2006})}\BibitemShut {NoStop}%
\bibitem [{\citenamefont {Ebbinghaus}\ \emph {et~al.}(2007)\citenamefont
  {Ebbinghaus}, \citenamefont {Kim}, \citenamefont {Heyden}, \citenamefont
  {Yu}, \citenamefont {Heugen}, \citenamefont {Gruebele}, \citenamefont
  {Leitner},\ and\ \citenamefont {Havenith}}]{Ebbinghaus07}%
  \BibitemOpen
  \bibfield  {author} {\bibinfo {author} {\bibfnamefont {S.}~\bibnamefont
  {Ebbinghaus}}, \bibinfo {author} {\bibfnamefont {S.-J.}\ \bibnamefont {Kim}},
  \bibinfo {author} {\bibfnamefont {M.}~\bibnamefont {Heyden}}, \bibinfo
  {author} {\bibfnamefont {X.}~\bibnamefont {Yu}}, \bibinfo {author}
  {\bibfnamefont {U.}~\bibnamefont {Heugen}}, \bibinfo {author} {\bibfnamefont
  {M.}~\bibnamefont {Gruebele}}, \bibinfo {author} {\bibfnamefont {D.-M.}\
  \bibnamefont {Leitner}}, \ and\ \bibinfo {author} {\bibfnamefont
  {M.}~\bibnamefont {Havenith}},\ }\href@noop {} {\bibfield  {journal}
  {\bibinfo  {journal} {Proc. Natl. Acad. Sci. U.S.A.}\ }\textbf {\bibinfo
  {volume} {104}},\ \bibinfo {pages} {20749} (\bibinfo {year}
  {2007})}\BibitemShut {NoStop}%
\bibitem [{\citenamefont {Tielrooij}\ \emph {et~al.}(2009)\citenamefont
  {Tielrooij}, \citenamefont {Timmer}, \citenamefont {Bakker},\ and\
  \citenamefont {Bonn}}]{Tielrooij2009}%
  \BibitemOpen
  \bibfield  {author} {\bibinfo {author} {\bibfnamefont {K.~J.}\ \bibnamefont
  {Tielrooij}}, \bibinfo {author} {\bibfnamefont {R.~L.~A.}\ \bibnamefont
  {Timmer}}, \bibinfo {author} {\bibfnamefont {H.~J.}\ \bibnamefont {Bakker}},
  \ and\ \bibinfo {author} {\bibfnamefont {M.}~\bibnamefont {Bonn}},\
  }\href@noop {} {\bibfield  {journal} {\bibinfo  {journal} {Phys. Rev. Lett.}\
  }\textbf {\bibinfo {volume} {102}},\ \bibinfo {pages} {198303} (\bibinfo
  {year} {2009})}\BibitemShut {NoStop}%
\bibitem [{\citenamefont {Heyden}\ \emph {et~al.}(2010)\citenamefont {Heyden},
  \citenamefont {Sun}, \citenamefont {Funkner}, \citenamefont {Mathias},
  \citenamefont {Forbert}, \citenamefont {Havenith},\ and\ \citenamefont
  {Marx}}]{Heyden2010}%
  \BibitemOpen
  \bibfield  {author} {\bibinfo {author} {\bibfnamefont {M.}~\bibnamefont
  {Heyden}}, \bibinfo {author} {\bibfnamefont {J.}~\bibnamefont {Sun}},
  \bibinfo {author} {\bibfnamefont {S.}~\bibnamefont {Funkner}}, \bibinfo
  {author} {\bibfnamefont {G.}~\bibnamefont {Mathias}}, \bibinfo {author}
  {\bibfnamefont {H.}~\bibnamefont {Forbert}}, \bibinfo {author} {\bibfnamefont
  {M.}~\bibnamefont {Havenith}}, \ and\ \bibinfo {author} {\bibfnamefont
  {D.}~\bibnamefont {Marx}},\ }\href@noop {} {\bibfield  {journal} {\bibinfo
  {journal} {Proc. Natl. Acad. Sci. U.S.A.}\ }\textbf {\bibinfo {volume}
  {107}},\ \bibinfo {pages} {12068} (\bibinfo {year} {2010})}\BibitemShut
  {NoStop}%
\bibitem [{\citenamefont {\'Smiechowski}\ \emph {et~al.}(2013)\citenamefont
  {\'Smiechowski}, \citenamefont {Forbert},\ and\ \citenamefont
  {Marx}}]{Smiechowski2013}%
  \BibitemOpen
  \bibfield  {author} {\bibinfo {author} {\bibfnamefont {M.}~\bibnamefont
  {\'Smiechowski}}, \bibinfo {author} {\bibfnamefont {H.}~\bibnamefont
  {Forbert}}, \ and\ \bibinfo {author} {\bibfnamefont {D.}~\bibnamefont
  {Marx}},\ }\href@noop {} {\bibfield  {journal} {\bibinfo  {journal} {J. Chem.
  Phys.}\ }\textbf {\bibinfo {volume} {139}},\ \bibinfo {pages} {014506}
  (\bibinfo {year} {2013})}\BibitemShut {NoStop}%
\bibitem [{\citenamefont {\'Smiechowski}\ \emph {et~al.}(2015)\citenamefont
  {\'Smiechowski}, \citenamefont {Sun}, \citenamefont {Forbert},\ and\
  \citenamefont {Marx}}]{Smiechowski2015}%
  \BibitemOpen
  \bibfield  {author} {\bibinfo {author} {\bibfnamefont {M.}~\bibnamefont
  {\'Smiechowski}}, \bibinfo {author} {\bibfnamefont {J.}~\bibnamefont {Sun}},
  \bibinfo {author} {\bibfnamefont {H.}~\bibnamefont {Forbert}}, \ and\
  \bibinfo {author} {\bibfnamefont {D.}~\bibnamefont {Marx}},\ }\href@noop {}
  {\bibfield  {journal} {\bibinfo  {journal} {Phys. Chem. Chem. Phys.}\
  }\textbf {\bibinfo {volume} {17}},\ \bibinfo {pages} {8323} (\bibinfo {year}
  {2015})}\BibitemShut {NoStop}%
\bibitem [{\citenamefont {Sun}\ \emph {et~al.}(2014)\citenamefont {Sun},
  \citenamefont {Niehues}, \citenamefont {Forbert}, \citenamefont {Decka},
  \citenamefont {Schwaab}, \citenamefont {Marx},\ and\ \citenamefont
  {Havenith}}]{Sun2014}%
  \BibitemOpen
  \bibfield  {author} {\bibinfo {author} {\bibfnamefont {J.}~\bibnamefont
  {Sun}}, \bibinfo {author} {\bibfnamefont {G.}~\bibnamefont {Niehues}},
  \bibinfo {author} {\bibfnamefont {H.}~\bibnamefont {Forbert}}, \bibinfo
  {author} {\bibfnamefont {D.}~\bibnamefont {Decka}}, \bibinfo {author}
  {\bibfnamefont {G.}~\bibnamefont {Schwaab}}, \bibinfo {author} {\bibfnamefont
  {D.}~\bibnamefont {Marx}}, \ and\ \bibinfo {author} {\bibfnamefont
  {M.}~\bibnamefont {Havenith}},\ }\href@noop {} {\bibfield  {journal}
  {\bibinfo  {journal} {J. Am. Chem. Soc.}\ }\textbf {\bibinfo {volume}
  {136}},\ \bibinfo {pages} {5031} (\bibinfo {year} {2014})}\BibitemShut
  {NoStop}%
\bibitem [{Note1()}]{Note1}%
  \BibitemOpen
  \bibinfo {note} {See Supplemental Material, \protect
  which includes Refs.~\protect \rev@citealp
  {Toukan1985,Lyubartsev2000,Raabe2007,Luzar1996,Kumar2007,Ramirez2004,%
  Ivanov2013,Heyden2012,Jonchiere2011,Skaf2000,Bursulaya1999b},
  for additional data, analyses, and computational details}\BibitemShut
  {NoStop}%
\bibitem [{\citenamefont {Toukan}\ and\ \citenamefont
  {Rahman}(1985)}]{Toukan1985}%
  \BibitemOpen
  \bibfield  {author} {\bibinfo {author} {\bibfnamefont {K.}~\bibnamefont
  {Toukan}}\ and\ \bibinfo {author} {\bibfnamefont {A.}~\bibnamefont
  {Rahman}},\ }\href@noop {} {\bibfield  {journal} {\bibinfo  {journal} {Phys.
  Rev. B}\ }\textbf {\bibinfo {volume} {31}},\ \bibinfo {pages} {2643}
  (\bibinfo {year} {1985})}\BibitemShut {NoStop}%
\bibitem [{\citenamefont {Lyubartsev}\ and\ \citenamefont
  {Laaksonen}(2000)}]{Lyubartsev2000}%
  \BibitemOpen
  \bibfield  {author} {\bibinfo {author} {\bibfnamefont {A.~P.}\ \bibnamefont
  {Lyubartsev}}\ and\ \bibinfo {author} {\bibfnamefont {A.}~\bibnamefont
  {Laaksonen}},\ }\href@noop {} {\bibfield  {journal} {\bibinfo  {journal}
  {Comp. Phys. Commun.}\ }\textbf {\bibinfo {volume} {128}},\ \bibinfo {pages}
  {565} (\bibinfo {year} {2000})}\BibitemShut {NoStop}%
\bibitem [{\citenamefont {Raabe}\ and\ \citenamefont
  {Sadus}(2007)}]{Raabe2007}%
  \BibitemOpen
  \bibfield  {author} {\bibinfo {author} {\bibfnamefont {G.}~\bibnamefont
  {Raabe}}\ and\ \bibinfo {author} {\bibfnamefont {R.~J.}\ \bibnamefont
  {Sadus}},\ }\href@noop {} {\bibfield  {journal} {\bibinfo  {journal} {J.
  Chem. Phys.}\ }\textbf {\bibinfo {volume} {126}},\ \bibinfo {pages} {044701}
  (\bibinfo {year} {2007})}\BibitemShut {NoStop}%
\bibitem [{\citenamefont {Luzar}\ and\ \citenamefont
  {Chandler}(1996)}]{Luzar1996}%
  \BibitemOpen
  \bibfield  {author} {\bibinfo {author} {\bibfnamefont {A.}~\bibnamefont
  {Luzar}}\ and\ \bibinfo {author} {\bibfnamefont {D.}~\bibnamefont
  {Chandler}},\ }\href@noop {} {\bibfield  {journal} {\bibinfo  {journal}
  {Phys. Rev. Lett.}\ }\textbf {\bibinfo {volume} {76}},\ \bibinfo {pages}
  {928} (\bibinfo {year} {1996})}\BibitemShut {NoStop}%
\bibitem [{\citenamefont {Kumar}\ \emph {et~al.}(2007)\citenamefont {Kumar},
  \citenamefont {Schmidt},\ and\ \citenamefont {Skinner}}]{Kumar2007}%
  \BibitemOpen
  \bibfield  {author} {\bibinfo {author} {\bibfnamefont {R.}~\bibnamefont
  {Kumar}}, \bibinfo {author} {\bibfnamefont {J.~R.}\ \bibnamefont {Schmidt}},
  \ and\ \bibinfo {author} {\bibfnamefont {J.~L.}\ \bibnamefont {Skinner}},\
  }\href@noop {} {\bibfield  {journal} {\bibinfo  {journal} {J. Chem. Phys.}\
  }\textbf {\bibinfo {volume} {126}},\ \bibinfo {pages} {204107} (\bibinfo
  {year} {2007})}\BibitemShut {NoStop}%
\bibitem [{\citenamefont {Ram\'\i{}rez}\ \emph {et~al.}(2004)\citenamefont
  {Ram\'\i{}rez}, \citenamefont {L\'opez-Ciudad}, \citenamefont {Kumar},\ and\
  \citenamefont {Marx}}]{Ramirez2004}%
  \BibitemOpen
  \bibfield  {author} {\bibinfo {author} {\bibfnamefont {R.}~\bibnamefont
  {Ram\'\i{}rez}}, \bibinfo {author} {\bibfnamefont {T.}~\bibnamefont
  {L\'opez-Ciudad}}, \bibinfo {author} {\bibfnamefont {P.}~\bibnamefont
  {Kumar}}, \ and\ \bibinfo {author} {\bibfnamefont {D.}~\bibnamefont {Marx}},\
  }\href@noop {} {\bibfield  {journal} {\bibinfo  {journal} {J. Chem. Phys.}\
  }\textbf {\bibinfo {volume} {121}},\ \bibinfo {pages} {3973} (\bibinfo {year}
  {2004})}\BibitemShut {NoStop}%
\bibitem [{\citenamefont {Ivanov}\ \emph {et~al.}(2013)\citenamefont {Ivanov},
  \citenamefont {Witt},\ and\ \citenamefont {Marx}}]{Ivanov2013}%
  \BibitemOpen
  \bibfield  {author} {\bibinfo {author} {\bibfnamefont {S.~D.}\ \bibnamefont
  {Ivanov}}, \bibinfo {author} {\bibfnamefont {A.}~\bibnamefont {Witt}}, \ and\
  \bibinfo {author} {\bibfnamefont {D.}~\bibnamefont {Marx}},\ }\href@noop {}
  {\bibfield  {journal} {\bibinfo  {journal} {Phys. Chem. Chem. Phys.}\
  }\textbf {\bibinfo {volume} {15}},\ \bibinfo {pages} {10270} (\bibinfo {year}
  {2013})}\BibitemShut {NoStop}%
\bibitem [{\citenamefont {Heyden}\ \emph {et~al.}(2012)\citenamefont {Heyden},
  \citenamefont {Sun}, \citenamefont {Forbert}, \citenamefont {Mathias},
  \citenamefont {Havenith},\ and\ \citenamefont {Marx}}]{Heyden2012}%
  \BibitemOpen
  \bibfield  {author} {\bibinfo {author} {\bibfnamefont {M.}~\bibnamefont
  {Heyden}}, \bibinfo {author} {\bibfnamefont {J.}~\bibnamefont {Sun}},
  \bibinfo {author} {\bibfnamefont {H.}~\bibnamefont {Forbert}}, \bibinfo
  {author} {\bibfnamefont {G.}~\bibnamefont {Mathias}}, \bibinfo {author}
  {\bibfnamefont {M.}~\bibnamefont {Havenith}}, \ and\ \bibinfo {author}
  {\bibfnamefont {D.}~\bibnamefont {Marx}},\ }\href@noop {} {\bibfield
  {journal} {\bibinfo  {journal} {J. Phys. Chem. Lett.}\ }\textbf {\bibinfo
  {volume} {3}},\ \bibinfo {pages} {2135} (\bibinfo {year} {2012})}\BibitemShut
  {NoStop}%
\bibitem [{\citenamefont {Jonchiere}\ \emph {et~al.}(2011)\citenamefont
  {Jonchiere}, \citenamefont {Seitsonen}, \citenamefont {Ferlat}, \citenamefont
  {Saitta},\ and\ \citenamefont {Vuilleumier}}]{Jonchiere2011}%
  \BibitemOpen
  \bibfield  {author} {\bibinfo {author} {\bibfnamefont {R.}~\bibnamefont
  {Jonchiere}}, \bibinfo {author} {\bibfnamefont {A.~P.}\ \bibnamefont
  {Seitsonen}}, \bibinfo {author} {\bibfnamefont {G.}~\bibnamefont {Ferlat}},
  \bibinfo {author} {\bibfnamefont {A.~M.}\ \bibnamefont {Saitta}}, \ and\
  \bibinfo {author} {\bibfnamefont {R.}~\bibnamefont {Vuilleumier}},\
  }\href@noop {} {\bibfield  {journal} {\bibinfo  {journal} {J. Chem. Phys.}\
  }\textbf {\bibinfo {volume} {135}},\ \bibinfo {pages} {154503} (\bibinfo
  {year} {2011})}\BibitemShut {NoStop}%
\bibitem [{\citenamefont {Skaf}\ and\ \citenamefont {Laria}(2000)}]{Skaf2000}%
  \BibitemOpen
  \bibfield  {author} {\bibinfo {author} {\bibfnamefont {M.~S.}\ \bibnamefont
  {Skaf}}\ and\ \bibinfo {author} {\bibfnamefont {D.}~\bibnamefont {Laria}},\
  }\href@noop {} {\bibfield  {journal} {\bibinfo  {journal} {J. Chem. Phys.}\
  }\textbf {\bibinfo {volume} {113}},\ \bibinfo {pages} {3499} (\bibinfo {year}
  {2000})}\BibitemShut {NoStop}%
\bibitem [{\citenamefont {Bursulaya}\ and\ \citenamefont
  {Kim}(1999)}]{Bursulaya1999b}%
  \BibitemOpen
  \bibfield  {author} {\bibinfo {author} {\bibfnamefont {B.~D.}\ \bibnamefont
  {Bursulaya}}\ and\ \bibinfo {author} {\bibfnamefont {H.~J.}\ \bibnamefont
  {Kim}},\ }\href@noop {} {\bibfield  {journal} {\bibinfo  {journal} {J. Chem.
  Phys.}\ }\textbf {\bibinfo {volume} {110}},\ \bibinfo {pages} {9656}
  (\bibinfo {year} {1999})}\BibitemShut {NoStop}%
\bibitem [{\citenamefont {Bellissent-Funel}\ \emph {et~al.}(1997)\citenamefont
  {Bellissent-Funel}, \citenamefont {Tassaing}, \citenamefont {Zhao},
  \citenamefont {Beysens}, \citenamefont {Guillot},\ and\ \citenamefont
  {Guissani}}]{Funel1997}%
  \BibitemOpen
  \bibfield  {author} {\bibinfo {author} {\bibfnamefont {M.-C.}\ \bibnamefont
  {Bellissent-Funel}}, \bibinfo {author} {\bibfnamefont {T.}~\bibnamefont
  {Tassaing}}, \bibinfo {author} {\bibfnamefont {H.}~\bibnamefont {Zhao}},
  \bibinfo {author} {\bibfnamefont {D.}~\bibnamefont {Beysens}}, \bibinfo
  {author} {\bibfnamefont {B.}~\bibnamefont {Guillot}}, \ and\ \bibinfo
  {author} {\bibfnamefont {Y.}~\bibnamefont {Guissani}},\ }\href@noop {}
  {\bibfield  {journal} {\bibinfo  {journal} {J. Chem. Phys.}\ }\textbf
  {\bibinfo {volume} {107}},\ \bibinfo {pages} {2942} (\bibinfo {year}
  {1997})}\BibitemShut {NoStop}%
\bibitem [{\citenamefont {Churakov}\ and\ \citenamefont
  {Kalinichev}(1999)}]{Churakov1999}%
  \BibitemOpen
  \bibfield  {author} {\bibinfo {author} {\bibfnamefont {S.~V.}\ \bibnamefont
  {Churakov}}\ and\ \bibinfo {author} {\bibfnamefont {A.~G.}\ \bibnamefont
  {Kalinichev}},\ }\href@noop {} {\bibfield  {journal} {\bibinfo  {journal} {J.
  Struct. Chem.}\ }\textbf {\bibinfo {volume} {40}},\ \bibinfo {pages} {548}
  (\bibinfo {year} {1999})}\BibitemShut {NoStop}%
\bibitem [{\citenamefont {Kalinichev}\ and\ \citenamefont
  {Bass}(1994)}]{Kalinichev1994}%
  \BibitemOpen
  \bibfield  {author} {\bibinfo {author} {\bibfnamefont {A.~G.}\ \bibnamefont
  {Kalinichev}}\ and\ \bibinfo {author} {\bibfnamefont {J.~D.}\ \bibnamefont
  {Bass}},\ }\href@noop {} {\bibfield  {journal} {\bibinfo  {journal} {Chem.
  Phys. Lett.}\ }\textbf {\bibinfo {volume} {231}},\ \bibinfo {pages} {301}
  (\bibinfo {year} {1994})}\BibitemShut {NoStop}%
\bibitem [{\citenamefont {Ma}\ and\ \citenamefont {Ma}(2011)}]{Ma2011}%
  \BibitemOpen
  \bibfield  {author} {\bibinfo {author} {\bibfnamefont {H.}~\bibnamefont
  {Ma}}\ and\ \bibinfo {author} {\bibfnamefont {J.}~\bibnamefont {Ma}},\
  }\href@noop {} {\bibfield  {journal} {\bibinfo  {journal} {J. Chem. Phys.}\
  }\textbf {\bibinfo {volume} {135}},\ \bibinfo {pages} {054504} (\bibinfo
  {year} {2011})}\BibitemShut {NoStop}%
\bibitem [{\citenamefont {Torii}(2011)}]{Torii2011}%
  \BibitemOpen
  \bibfield  {author} {\bibinfo {author} {\bibfnamefont {H.}~\bibnamefont
  {Torii}},\ }\href@noop {} {\bibfield  {journal} {\bibinfo  {journal} {J.
  Phys. Chem. B}\ }\textbf {\bibinfo {volume} {115}},\ \bibinfo {pages} {6636}
  (\bibinfo {year} {2011})}\BibitemShut {NoStop}%
\bibitem [{\citenamefont {Torii}(2014)}]{Torii2014}%
  \BibitemOpen
  \bibfield  {author} {\bibinfo {author} {\bibfnamefont {H.}~\bibnamefont
  {Torii}},\ }\href@noop {} {\bibfield  {journal} {\bibinfo  {journal} {J.
  Chem. Theory Comput.}\ }\textbf {\bibinfo {volume} {10}},\ \bibinfo {pages}
  {1219} (\bibinfo {year} {2014})}\BibitemShut {NoStop}%
\bibitem [{\citenamefont {Segr{\`e}}\ \emph {et~al.}(1997)\citenamefont
  {Segr{\`e}}, \citenamefont {Herbolzheimer},\ and\ \citenamefont
  {Chaikin}}]{Segre1997}%
  \BibitemOpen
  \bibfield  {author} {\bibinfo {author} {\bibfnamefont {P.~N.}\ \bibnamefont
  {Segr{\`e}}}, \bibinfo {author} {\bibfnamefont {E.}~\bibnamefont
  {Herbolzheimer}}, \ and\ \bibinfo {author} {\bibfnamefont {P.~M.}\
  \bibnamefont {Chaikin}},\ }\href@noop {} {\bibfield  {journal} {\bibinfo
  {journal} {Phys. Rev. Lett.}\ }\textbf {\bibinfo {volume} {79}},\ \bibinfo
  {pages} {2574} (\bibinfo {year} {1997})}\BibitemShut {NoStop}%
\end{thebibliography}
\end{document}